\documentstyle[12pt]{article}
\def\be{\begin{equation}}
\def\ee{\end{equation}}
\def\D{{\cal D}}
\def\C{{\cal C}}
\def\G{{\cal G}}
\def\Im{{\rm Im}}
\def\Re{{\rm Re}}
\def\e{\epsilon}
\def\t{\tau}
\def\det{{\rm det}}
\def\d{\partial}
\def\S{\Sigma}
\def\L{\Lambda}

\begin{document}
\baselineskip=15pt
\title{2+1 Sector of 3+1 Gravity}
\author{Jerzy Lewandowski and Jacek Wi\'sniewski}
\maketitle
\centerline{\it Instytut Fizyki Teoretycznej, Wydzia\l \ Fizyki}
\centerline{\it Uniwersytet Warszawski, ul Ho\.za 69, 00-681, Warsaw, Poland}

\begin{abstract} The rank-2 sector of classical $3+1$ dimensional 
Ashtekar gravity is considered. It is found that the consistency of 
the evolution equations with the reality of the volume requires that 
the 3-surface of initial data is foliated by 2-surfaces tangent
to degenerate triads. In turn, the degeneracy is preserved
by the evolution. The 2-surfaces 
behave like $2+1$ dimensional empty spacetimes with a massless 
complex field propagating along each of them. The results
provide  some evidence for the issue of evolving a non-degenerate
gravitational field  into a degenerate sector of Ashtekar's
phase space. 
\end{abstract}

\section{Introduction.}
One of the fundamental aspects of the general relativity
is which  configuration should be interpreted as an absence
of  a gravitational field. Obviously, a flat metric is not
the right choice. It breakes the diffeomorphism symmetry,
tells the particles which curves are straight lines
and which are not. From the point of view of  traditional
gravity, the only promising candidates are either $g_{\mu\nu}=0$
or $g^{\mu\nu}=0$. However, neither of them exists 
in the space of the solutions of the Einstein equations naturally.  
Ashtekar's  phase space for 3+1 gravity \cite{ash}, on the other hand, 
containes the zero data for which both the Ashtekar connection
and densitised triad vanish.  It belongs to
the  set of data which can not be used to construct  
non-degenerate 3+1 metric tensors. This set is often called 
the degenerate
sector of the 3+1 theory and its relevance is justified
by the relevance of the most degenerate, zero data. 
Various aspects of the degenerate
sector were studied for instance by Jacobson and Romano \cite{jacrom}, 
Bengtsson \cite{ben1} and Reisenberger \cite{reis}.
Matschull \cite{mat} showed that also degenerate Ashtekar's data 
provides  the corresponding 4 dimensional manifold      
with a causal structure. For many relativists this is yet another 
argument sufficient to learn more about that sector.
The main motivation for our work comes from the recent paper
of Jacobson \cite{jac}.  He considered a general case, when the Ashtekar data
defines on an initial 3-surface $\S$ a triad of colinear
vector densities (instead of three linearly independent ones),
which he  called `1+1 sector of  3+1 gravity'.  Jacobson
solved the Einstein equations completely for that case. A general
solution has the following  features. The degeneracy is invariant in time
and the integral curves defined by the triad
and foliating $\S$ behave like `1+1 dimensional vacuum spacetimes with a pair
of massless complex valued   fields propagating along them'. 
\bigskip

Below, we solve the Ashtekar-Einstein
equations for a data of a lower degeneracy, where  
triad defines at a point two linearly independent vectors.
The preservation of the reality by the evolution
implies the existence of a foliation of $\S$ into the  
integral 2-surfaces tangent to a given triad.
That integrability condition weakly commutes with the hamiltonian
which ensures that the degeneracy is preserved.  
Analogously to the Jacobson's case, the Ashtekar-Einstein equations
of the 3+1 dimensional gravity make the 2-surfaces behave
like 2+1 dimensional empty spacetimes with an extra massless complex
field assigned to each surface and progatating along it.

In  the last section we  indicate  relevance of Jacobson's and 
the current results for some issues of the evolution of
a non-degenerate gravitational field, characterise the remaining
degeneraces and briefly discuss a corresponding degenerate case in 
the Ashtekar theory applied to the Euclidean gravity.       

Some of our results have been rederived  in a quite elegant way
by Matschull in terms of his quasi covariant approach \cite{mat2}.  
\bigskip

\noindent{\bf Ashtekar's theory.}
The Ashtekar theory is a canonical theory on a space-time
manifold $\Sigma\times R$ where $\Sigma$ is a theree-real-surface
of initial data (the `space') and  $R$ is the one dimensional  space 
of values for a time parameter. The phase space consists of the pairs
of fields $(A, E)$ where $A$ is an algebra $sl(2,C)$ valued one-form on
$\Sigma$ and $E$ is an $sl(2,C)$ valued vector density of the weight 1
defined on $\Sigma$. Using  local coordinates $(x^a)= (x^1,x^2,x^3)$ on 
$\Sigma$  and a basis $(\tau_i)=(\tau_1,\tau_2,\tau_3)$ of $sl(2,C)$
we write
\be
A\ =\ A^i_a\tau_i\otimes dx^a, \ \ E\ =\ E^{ia}\tau_i\otimes \partial_a,
\ee      
where $A^i_a$ and $E^{ia}$ are complex valued functions defined on $\Sigma$.
In $sl(2,C)$ we fix the standard bilinear complex valued inner product
\be
               k(v,w)\ =\ -2 {\rm tr} (vw).
\ee
The variables $(A,E)$ are canonically conjugate in the sence
of the following, the only nonvanishing Poisson bracket
\be \label{poisson}
\{A^i_a(x), E^{jb}(y)\}\ =\ i k^{ij} \delta^b_a \delta(x,y).
\ee

A data $(A,E)$ is accompanied by  Lagrange multipliers,
a $-1$ weight density $N$ (the densitiesed laps function), a vector field
$N^a$ (the familiar shift) and an $sl(2,C)$ valued function $\Lambda$, all 
defined on $\Sigma$. The hamiltonian is given by 
\be
H\ =\ \C_N\ +\ \C_{\vec N}\ +\ \G_{\Lambda}
\ee
\be
\C_N\ :=\ \int_\Sigma d^3x N\C(A,E)\ :=\ -{1\over 2}
\int_\Sigma d^3x N F_{ab}^i E^{ja}E^{kb}c_{ijk},  
\ee
\be
C_{\vec N} \ :=\ \int_\Sigma N^a \C_a(A,E)\ :=\ -i\int_\Sigma d^3x 
N^a F^{i}_{ab}E^{b}_i,
\ee
\be
\G_{\Lambda}\ :=\ \int_\Sigma d^3x \Lambda_i \G^i(A,E) \ 
:=\ i\int_\Sigma d^3x \Lambda_i\D_a E^{ia},
\ee
where 
\be
F := {1\over 2} F_{ab}^i \tau_i\otimes dx^a\wedge dx^b\ :=\ dA\ +\ A\wedge A,
\ee
is the curvatute of $A$ and $\D_a$ is the standard covariant derivative
\be
\D_a w^i \ := \partial_a w^i + c^i_{jk}A_a^iw^k
\ee

(which applied to $E^{ia}$ ignores the $a$ index).

The constraints $\C_N,\,  \C_{\vec{N}},\, \G_{\L}$ generate
the time evolution, diffeomorphisms in of $\S$ and the Yang-Mills
gauge transformations, the last being $(A,E)\mapsto (g^{-1}Ag+g^{-1}dg,\,
g^{-1}Eg)$ where $g$ is any $SL(2,C)$ valued function.  
\medskip

Apart from the resulting constraint equations, the data $(A,E)$
is subject to the following  reality conditions   
\be \label{real}
\Im \big(E^{ia}E^b_i\big)\ =\ 0,\ \ \Im \big(\{E^{ia}E_i^b,\, \C_N\}\big)\ 
=\ 0.
\ee   
As long as the matrix 
$(E^{ia})_{i,a=1,...3}$ 
is of the rank 3 and 
the signature of the symmetric matrix $(E^{ia}E_i^b)_{a,b=1,...,3}$ is
$(+++)$ one constructs an ADM data from $(A,E)$ and the Ashtekar
theory is equivalent to the Einstein gravity with the Lorentzian signature. 
However, the theory naturally extends to degenerate cases, when
the ranks are lower then 3.  

\bigskip
\section{The main result.}
We will be concerned now with the Ashtekar's equations
for a degenerate case characterised below. (All the 
considerations are local.) 
\bigskip

\noindent{\bf The rank 2 assumption.} 
What we are assuming in this section is the following degeneracy condition
to be satisfied everywhere on $\Sigma$ at the instant $t=0$,
\be \label{rank}
{\rm rank}(E^{ia})\ =\ 2,\ \ {\rm and}\ \  
{\rm signature}(E^{ia}E_i^b)\ =\ (++0).
\ee 
(See the next section for the list of the remaining cases.)
In the non-degenerate case, the second matrix in (\ref{rank}) is related 
to the corresponding metric tensor on $\S$ via $E^{ia}E_i^b=\sqrt{q} q^{ab}$.
However,
that matrix has also its own geometric interpretation. In a non-degenerate
case, the 2-area element of a 2-surface $x^a(r,s)$ parametrised by $(r,s)$, 
is $\sqrt{E_i^aE^{ib}f_af_b} dr\wedge ds$, where $f_a:=\e_{abc}x^bx^c$. This 
interpretation, sometimes emphasised in  lectures on the Ashtekar variables
(see for instance \cite{giu}) extends naturally to the degenerate sector.  

We fix in $sl(2,C)$ an orthonormal basis $(\tau_i)$ such
that 
\be
[\tau_i,\tau_j]\ =\ \epsilon_{ijk}\tau_k.
\ee
It follows from the first of the reality conditions (\ref{real})
and from the above assumptions that there exists an $SL(2,C)$
valued function $g$  and two real, linearly
independent vector field densities $e_1$ and $e_2$ defined on $\Sigma$ 
such that  
\be \label{gauge}
g^{-1}Eg\ =\  \tau_1\otimes e_1\ +\ \tau_2\otimes e_2.
\ee 
This observation will be usefull below.

\bigskip

\noindent{\bf Existence of a foliation tangent to $E^{ia}\partial_a$.} 
The reality of the vectors $e_1$ and $e_2$ in (\ref{gauge}) shows that
at each point of $\Sigma$, $E$ defines a real, 2-dimensional subspace
of the tangent space which is invarianat upon the Yang-Mills gauge 
transformations. We will see that the resulting family of 
vector subspaces is integrable.    
Any given $E^{ia}$, denote by ${\rm det}E$ the determinant of the matrix
$(E^{ia})_{i,a=1,2,3}$. Since $E$ is degenerate at $t=0$, 
the determinant vanishes. However the Poisson bracket $\{{\rm det}E,H\}$
does not vanish in general. The reality conditions (\ref{real}) imply
that  
\be 
\Im \big(\{{\rm det}E,H\}\big)\ =\ 0.
\ee
The determinant is invariant with respect to the gauge transformations
$E\rightarrow g^{-1}Eg$ so we can use (\ref{gauge}) to evaluate the 
Poisson bracket, and derive
\be \label{det}
\{{\rm det}E,H\}\ =\ i\epsilon_{abc}[e_1,\, e_2]^ae_1^be_2^c.
\ee
That is, $E$ defines a foliation of $\Sigma$ into 2-sub-surfaces
tangent to  all the vector fields $\Im (E^{ia})\d_a$ and 
$\Re(E^{ia})\d_a$, $i=1,2,3$. 

\bigskip
\noindent{\bf Propagation of the degeneracy and of the integrability.} 
From now on let $(x^a)$ be coordinates such that 
\be \label{int}
E^{i3}\ =\ 0, \ \ i=1,2,3,
\ee 
(at $t=0$). The functional $E^{i3}$ considered as a constraint
commutes weakly with the hamiltonian, namely
\be\label{weak}
\{E^{i3}, \C_N\}\ =\ i\e_{ijk} E^{jb}(\d_bE^{k3} + A^m_bE^{n3}\e_{kmn})
+ \e_{ijk} \G^jE^{3k}.
\ee
Therefore, it should be preserved by the  hamiltonian evolution, that is
\be \label{evolEt}
E^{i3}(t)\ =\ 0
\ee
in the coordinates $(x^a,t)$. 

\bigskip
\noindent{\bf Spliting of $(A,E)$ into  the tangent and the 
transversal components.}
We will use the following index notation below, 
\be
(x^a)\ =\ (x^B, x^3),
\ee
that is $(x^B)$ parametrise the leaves of the foliation $x^3=const$.
It is convienient now to analyse the constraints and the evolution 
equations from the point of view of the spliting of $(A,E)$ into
the components $({}^2\!A:=A^i_B\t_i\otimes dx^B,\ {}^2\!E:=E^{jC}\t_j
\otimes \d_C)$ tangent to the leaves,
and the remaining transversal component $A^i_3$. 

Those of the constraints and the Poisson brackets $\{A^i_a,\C_N\}$ and
$\{E^{ia},\C_N\}$ which neither involve $A^i_3$ nor the transversal derivatives
$\d_3A^i_b$, $\d_3E^{iB}$ are 
\be\label{scalar}
\C(A,E)\ =\ -{1\over 2} \e_{ijk} F_{AB}^i E^{jA}E^{kB}\ =:\ 
\C^{(2)}({}^2\!A,{}^2\!E),
\ee
\be\label{diffeo}
\C_B(A,E)\ =\ -i F^i_{BC}E_i^C\ =:\ \C^{(2)}({}^2\!A,{}^2\!E),
\ee
\be \label{gauss}
\G^i(A,E)\ =\ i\D_B E^{iB}\ =: \G^{i(2)}({}^2\!A,{}^2\!E).
\ee
\medskip

\be\label{evolE}
\{E^{iB}, \C_N\}\ =\ i\e_{ijk} \D_C(E^{jC}NE^{kB}),
\ee
\be\label{evolA}
\{A^i_B,\C_N\}\ =\ i\e_{ijk} E^{jC}F^k_{CB}.
\ee
    
All the remaining constraints and the evolution Poisson brackets are
\be\label{diffeo3}
\C_3(A,E)\ =\ -iE^{B}_i (\d_3A^i_B\ -\ \D_B A^i_3),
\ee
\be\label{evolAt}
\{A^i_3, \C\}\ =\ i\e_{ijk}E^{jB}(\D_B A^k_3\ -\ \d_3A^k_B).
\ee 

\bigskip
\noindent{\bf The equations on $({}^2\!A,{}^2\!E)$.} The equations 
(\ref{scalar},\ref{diffeo},\ref{gauss}) and (\ref{evolA},\ref{evolE})
come down to a family of equations labelled by values of the coordinate
$x^3$. The fields subject to them are only $({}^2\!A,{}^2\!E)$. Obviously,
the components $F^i_{BC}$ of the curvature of $A$ set up
the curvature of ${}^2\!A$ defined on each leaf,
\be
{}^2\!d\ {}^2\!A\ +\ {}^2\!A\wedge {}^2\!A\ =\ {1\over 2} 
F^i_{BC}\t_i\otimes dx^B\wedge
dx^C.
\ee

Let us fix $x^3$ and study  the  2+1 dimensional theory 
of the data $({}^2\!A,{}^2\!E)$ restricted to the corresponding leaf
$\S_{x^3}$ resulting from the equations (\ref{scalar}, \ref{diffeo},
\ref{gauss}) and (\ref{evolA}, \ref{evolE}).
  
The Poisson bracket (\ref{poisson}) induces the following bracket
on $\S_{x^3}$,
\be\label{pois2}
\{A^i_B,E^{jC}\}^{(2)}\ =\ i k^{ij}\delta^C_B\delta^{(2)}(x,y).
\ee
Consider a hamiltonian 
\be\label{ham2}
H^{(2)}({}^2\!A,{}^2\!E,N,N^A,\L^i)\ :=\ \int_{\S_{x^3}}d^2x \big(
N \C^{(2)}\ +\ N^BC^{(2)}_B\ +\ \L_i\G^{(2)i}\big).
\ee
 It is easy to see that as long as we consider only the shifts
which preserve the leaves, that is such that 
\be
N^3\ =\ 0,
\ee
then 
\be
\{A^i_B,H\}\ =\ \{A^i_B,H^{(2)}\}^{(2)}, \ \ \{E^{iB}, H\}\ =\ 
\{E^{iB},H^{(2)}\}^{(2)}.
\ee
The hamiltonian $H^{(2)}$ algebraically  coincides with
the Ashtekar's hamiltonian for the $2+1$ vacuum gravity
\cite{ben2}.\footnote{An easy way to derive this fact is to start with
the Ashtekar action of the $3+1$ dimensional gravity
and next restrict it to the case when the spacetime is the
orthogonal product of a spacelike 1 dimensional manifold
with a 2+1 manifold.} 
The only difference is, that to define the  real 
$2+1$ gravity  the fields $(A^i_B, E^{jC})$
should satisfy the following reality condition for every 
space index $B$, 
\be\label{real2}
A^i_B\t_i,\ iE^{jB}\t_j\ \in\ so(2,1)
\ee 
where by $so(2,1)\subset sl(2,C)$ we mean the real Lie subalgebra
spanned by the generators $(i\t_1,i\t_2,\t_3)$,
\be
so(2,1)\ =\ {\rm span}(i\t_1,i\t_2,\t_3).
\ee  

\bigskip
\noindent{\bf The 3+1 theory reality conditions.}
Apart from the constraints $\C^{(2)},\ \C^{(2)}_B,$ and $\G^{(2)i}$
the fields $({}^2\!A,\, {}^2\!E)$ are restricted by the reality conditions
(\ref{real}). From the equation (\ref{gauge}) resulting from the first
of the reality  conditions, we conclude that there does exist 
a Yang-Mills gauge transformation $E\rightarrow g^{-1}Eg$ such that
(\ref{real2}) is obeyed by $E$. Let us restrict ourselves to
such gauge, that is assume that $E$ satisfies (\ref{real2}). 
Clearly, if  $A^i_B$ also happens to satisfy
(\ref{real2}) then $\Im \big(\{ E^{ia}E_i^{b},H\}\big)=0$.
Indeed, if we use $i\, {}^2\!E$ as a variable, then all the factors $i$
get absorbed in (\ref{evolE}, \ref{evolA}), and the only operations
used on the $so(2,1)$ valued fields are the derivatives 
and commutators. Thus
\be
\{iE^{iB}, \C_N\}\ \in so(2,1)
\ee
itself as long as $A^i_B$ satisfies (\ref{real2}). It is not
hard to show that $A^i_B\t_i\in so(2,1)$ is also a necessary
condition for the $3+1$ reality conditions (\ref{real}) to
hold, provided the Gauss constraint is satisfied. To see that 
momentarily, notice, that it follows from (\ref{gauge}) that 
there exists a Yang-Mills gauge transformation 
generated by some $so(2,1)$ valued function and coordinates 
$({x'}^A,x^3)$, such that (dropping the primes)
\be
E\ =\ e^\sigma(\t_1\d_1\ +\ \t_2\d_2),
\ee
$\sigma$ being a real function  defined on $\S$. It is enough to prove
that upon this gauge choice $A^i_B\t_i\in so(2,1)$. Now, 
$\D_BE^{iB}=0$ implies that
\be
\Im(A^3_B)\ =\ 0, \ \ {\rm and}\ \ A^1_2\ =\ A^2_1,
\ee  
whereas the reality conditions (\ref{real}) read 
\be
0\ =\ \Re (A^2_2)\ =\ \Re (A^1_1)\ = \Re (A^1_2),
\ee
which completes the proof that the one form ${}^2\!A$ takes values in
$so(2,1)$. 

Concludding, the reality conditions of the $3+1$ theory and the 
constraints $\C, \C_B, \G^i$ get reduced to the constraints 
$\C^{(2)}, \C^{(2)}_B, \G^{(2)i}$ and 
the reality conditions of the $2+1$ theory given by $({}^2\!A,{}^2\!E)$
on each leaf $\S_{x_3}$ separately. 
 Moreover, the evolution generated by the hamiltonian 
of the $3+1$ theory coincides with the evolution equations
of the vacuum $2+1$ gravity. Hence, we have solved completely the equations
given by (\ref{scalar},\ref{diffeo},\ref{gauss}) and (\ref{evolA},
\ref{evolE}).   
 
\bigskip
\noindent{\bf The equations for the field $A^i_3$.} Contrary to 
the equations solved above,  the equations
on $A^i_3$ (that is the remaining equations (\ref{diffeo3}, \ref{evolAt})
contain also the transversal derivatives $\d_3 A^i_B$. 
One might conclude, that there is some interaction between the leaves.
However, the scalar and diffeomorphism constraint equations $\C=0,\ \C_B=0$
(and the rank assumptions (\ref{rank})) imply that
\be
F^i_{BC}\ =\ 0.
\ee  
Therefore,  we can chose a Yang-Mills gauge in  such that
\be \label{gaugeA}
A^i_B\ =\ 0.
\ee
In such a gauge, all the transversal derivatives disappear
from the equations, and the remaining equations (\ref{diffeo3},
\ref{evolAt}) read (for a given $E^{iB}$ and unknown complex valued
$A^i_3$)
\be \label{diffeo30}
\d_B \big(A^i_3 E^{iB}\big)\ =\ 0,
\ee
\be\label{evolAt0}
{\d\over \d t}A^i_3\ =\ i\e_{ijk} E^{jB}\d_B A^k_3,
\ee
(if we use a basis $(i\t_1, i\t_2, \t_3)$ and multiply the first
equation by $i$ then  the coefficients become real.)
To make  that equations really simple, let us apply our knowledge about
the solutions $({}^2\!A,{}^2\!E)$ of the vacuum $2+1$ gravity equations. 
Since they are just given by a slicing of the 3 dimensional Minkowski
spacetime, there exists a choice of coordinates $({x'}^A,{x'}^3,t')$
in $\Sigma\times R$ (possibly defining a different slice 
${\Sigma'}_{t'=const}$) such that (dropping the primes)
\be\label{final1}
A^i_A\ =\ 0,\ \ E\ =\ (\t_1\d_1\ +\ \t_2 \d_2).
\ee    
 Then a general solution to the equations (\ref{diffeo3},\ref{evolAt})
is given by a complex valued potential $\lambda$ defined on $\S\times R$
which satisfies the wave equation
\be\label{final2}
\lambda_{,tt}\ -\ \lambda_{,11}\ -\ \lambda_{,22}\ =\ 0,
\ee 
via
\be\label{final3}
A^1_3\ =\ -\lambda_2,\ A^2_3\ =\ \lambda_1,\ A^3_3\ =\ -i\lambda_t.
\ee

\section{Conclusions.}  
\noindent{\bf Summary.} We have solved locally  the Ashtekar-Einstein
equations for a general case of a data $(A,E)$ which satisfies
the degeneracy conditions (\ref{rank}). 
The reality conditions of the $3+1$ dimensional theory with the Lorentzian
signature imply that $E(t=0)$ defines in $\Sigma$ a two dimensional 
foliation. That integrability property expressed by (\ref{int})
propagates due to the weak comutation relation (\ref{weak}). 
Locally, in an appropriate Yang-Mills gauge (\ref{gaugeA})  whose 
existence  is guarantied 
by the constraints the  theory of the sector (\ref{rank}) comes down
to a family of independent  theories leaving on the leaves of the foliation.
For every leaf, the corresponding theory is the vacuum $2+1$ dimensional
gravity (\ref{pois2},\,\ref{ham2}) (of the $(-++)$ signature) plus a 
 complex valued field $A^i_3$ subject to a constraint
equation (\ref{diffeo30}) and an evolution equation (\ref{evolAt0}).
A general $(A,E)$ may be described in the following way.
There exist coordinates $(x^A,x^3)$ on $\S$
such that (\ref{evolEt}) holds and, for each value of $x^3$, 
the components $\big(A^i_B, \,E^{iB},\, N, \, N^A, \L^i \big)(x^A,x^3,t)$ 
define on $\Sigma_{x^3=const}\times R$ a $2+1$ dimensional 
vaccuum gravitational
field via Ashtekar's anzatz. There exist coordinates
$({x'}^A,{x^3},t')$ 
in $\Sigma\times R$, such that 
all the Lagrange multipliers vanish and $(A,E)({x'}^A,{x^3},t')$
is given by (\ref{final1}-\ref{final3}).    

It has been natural to restrict the diffeomorphisms of $\S$ to
the ones preserving the leaves of the foliation, that
is such that  $N^3=0$. However, the choice 
$E^{i3}=0$ implies only $N^3=N^3(x^3)$. So there is still a gauge 
symetry which mixes  the theories living on different leaves.  

As indicated at the beginning, all our results are  local.

\bigskip
 
\noindent{\bf The other degenerate sectors.}
It is worth to list all possible kinds of degeneracy which
can potentially occure in Ashtekar's theory for the Lorentzian sygnature. 
If the rank of $(E^{ia})$ is maximal then so is the rank of the
`2-area matrix' $E^{a}_iE^{ib}$. However, since $E$ is complex valued, 
in general the rank of the 2-area matrix is lower equal the rank
of $(E^{ia})$. If we restrict ourselves to semi-positive 
definite case of the 2-area matrix, the possible cases are $(0,0)$,
$(1,0),\ (1,1),\ (2,1),\ (2,2),\ (3,3)$, where the numbers indicate
the ranks of the triad matrix and the 2-area matrix respectively.
The cases which are known thus far  are $(3,3)$, $(0,0)$, $(1,1)$ 
and $(2,2)$. An example of the $(1,0)$ case is 
$E=(\t_1+i\t_2)\otimes X$
and an example of the $(2,1)$ case is $E=(\t_1+i\t_2)\otimes X+
\t_3 Y$, where $X,Y$ are complex vector field densities on $\S$.
Some  relevance of these cases is indicated  below.     

\bigskip
\noindent{\bf Relevance for the issues of the time evolution of a classical
gravitational field.} 
Shortly after introducing the Ashtekar variables it was conjuctured
that may be the time evolution can carry a nondegenerate
data through a degenerate sector of Ashtekar's phase space
back to the non-degenerate sector (of either the Lorentzian or perhaps
Euclidean gravity). However no evidence
proving or disproving that conjucture has been provided.
The recent results on the $1+1$ sector \cite{jac} and the current work
show that a non-degenerate data of the Lorentzian gravity 
can not evolve  into  the rank  $(1,1)$ or $(2,2)$ sector. 

It is worth noting that the arguments used above do not continue
to be true in the Eulidean case.
\bigskip

\noindent{\bf Degenerate sectors of the Euclidean gravity: similarities
and differences.} 
The Ashtekar theory applies also to the Euclidean gravity. All
one has to do, is to remove the $i$ factor from the Poisson brackets,
and assume that all the fields are real. This reduces the number
of possible degeneracies to the cases $(0,0)$, $(1,1)$, $(2,2)$
and $(3,3)$. Let us discuss briefly the $(2,2)$ case.
However at each point of $\S$, $E$ defines again a 2-dimensional subspace
of the tangent space, the obtained distribution
may be non-integrable. Indeed, now there is no $i$ factor
in the right hand side of (\ref{det}).   So, locally, we have two cases:
either $(i)$ $E$ defines a foliation of $\S$ into 2-submanifolds 
(by an assumption), or $(ii)$ the commutators of  vector
fields  $E^{ia}\d_a$ generate at each point  a third, linearly 
independent vector field.
In the case $(i)$, the results of the Sec 2 summarised above
still apply modulo removing the $i$ from the Poisson brackets
and few changes of signes (this time we don't have to prove 
the reality, since everything is real by definition). 
 In the case $(ii)$, on the other hand, the calculation
analogous to that of eq. (\ref{det})  shows 
that ${\d\over \d t} \det E \not= 0$ at $t\not=0$. This proves, that
for a sufficiently small $t$, the evolution carries
our $(A,E)$ out of the $(2,2)$ sector  into a non-degenerate
data. 
\bigskip

\noindent{\bf Acknowledgements.}
We are grateful to Ted Jacobson for presenting to us his
results about the $1+1$ dimensional case before they were
written down and to Hans Matschull for confirming our calculations
in a quite neat way. We have benefited a lot from the discussions
with Peter Aichelburg, Abhay Ashtekar, Robert Budzy\'nski, 
Don Marolf, Krzysztof
Meissner,  Jose Mourao, Herman Nicolai, Thomas Thiemann,
Helmut Urbantke
and participants of the ESI workshop on Quantum Gravity
where this work was finished. JL thanks 
Peter Aichelburg and the Erwin Shr\"odinger Institute
for their warm hospitality. 
  This work was supported in part by  the KBN grant 2-P302 11207.

\end{document}